\documentclass{emulateapj}

\newcommand{\kms}{km~s$^{-1}$}
\newcommand{\ie}{{\it i.e., }}
\newcommand{\etal}{{\it et al.~}}
\newcommand{\eg}{{\it e.g., }} 
\newcommand{\msun}{M$_\odot$} 

\slugcomment{To appear in Astrophysical Journal} 

\shorttitle{Radio Emission from SNIa}
\shortauthors{Panagia, \etal}

\begin{document}

\title{A Search for Radio Emission from Type Ia Supernovae} 

\author{Nino Panagia\altaffilmark{1,2}, 
Schuyler D. Van Dyk\altaffilmark{3}, 
Kurt W. Weiler\altaffilmark{4}, 
Richard A. Sramek\altaffilmark{5}, \\
Christopher J. Stockdale\altaffilmark{6},
Kimberly P. Murata\altaffilmark{7}}
\altaffiltext{1}{Space Telescope Science Institute, 3700 San Martin
Drive, Baltimore, MD 21218;  panagia@stsci.edu}
\altaffiltext{2}{Also, Istituto Nazionale di Astrofisica (INAF), Rome,
Italy; and Supernova Ltd., Virgin Gorda, British Virgin Islands. }
\altaffiltext{3}{Spitzer Science Center, Caltech, Pasadena, CA 91125; 
 vandyk@ipac.caltech.edu}
\altaffiltext{4}{Naval Research Laboratory, Code 7213, Washington, DC
20375-5320; Kurt.Weiler@nrl.navy.mil}
\altaffiltext{5}{National Radio Astronomy Observatory, P.O.~Box 0,
Socorro, NM 87801; dsramek@nrao.edu}
\altaffiltext{6}{Department of Physics, Marquette University, P.O. Box
1881, Milwaukee, WI 53201-1881; christopher.stockdale@marquette.edu}
\altaffiltext{7}{9031 Ashmeade Drive, Fairfax, VA 22032; kpm1@cornell.edu}

\begin{abstract}
We present and discuss the radio observations of 27 Type Ia supernovae
(SNe Ia) observed over two decades with the Very Large Array.  No SN Ia
has been detected so far in the radio, implying a very low density for
any possible circumstellar material established by the progenitor, or
progenitor system, before explosion.  We derive $2\sigma$ upper limits
to a steady mass-loss rate for individual SN systems as low as $\sim 3
\times 10^{-8}$ M$_\odot$ yr$^{-1}$, discriminating strongly against
white dwarf accretion via a stellar wind from a massive binary companion
in the symbiotic star, an example of the ``single degenerate''
scenario.  However, a white dwarf accreting from a relatively low mass
companion via a sufficiently high efficiency ($>60-80$ \%), Roche lobe
overflow is still consistent with our limits. The ``double degenerate''
merger scenario also cannot be excluded.
\end{abstract}

\keywords{(stars:) supernovae: general, radio supernovae, stars: mass
loss, (stars:) white dwarfs, (stars:) binaries: symbiotic, (stars:)
binaries: close, radio continuum: stars}

\section{Introduction\label{intro}}

Supernovae (SNe) are among the most energetic events in the Universe.
Determining the properties of the progenitor stars or stellar systems
remains an important unsolved problem in astrophysics. Few SNe have had
their progenitor stars directly identified but the post-explosion radio
emission from SNe provides insight into the nature of the progenitor
stars and their last stages of evolution.  The phenomenon of radio SNe
(RSNe) has  been best modeled by synchrotron emission resulting from the
interaction of the SN shock with circumstellar material (CSM)
established by pre-SN mass-loss from the progenitor system before the
explosion; likely from the progenitor itself, or possibly from the
progenitor's companion in the case of a binary system (Chevalier 1982a,
1982b; Sramek et al. 1984, Weiler et al. 1986).

Although it has been generally accepted that Type Ia SNe (SNe Ia) result
from the thermonuclear explosion of a white dwarf (WD) star (e.g.,
Nomoto, Thielemann \& Yokoi 1984, Branch et al. 1985), two fundamental
questions remain: 1) Is the exploding WD of Chandrasekhar or
sub-Chandrasekhar mass?  2) If the former, how does the exploding WD
(typically $\sim 0.6\ M_{\odot}$; e.g., Ritter \& Burkert 1986)
accumulate enough mass to approach the Chandrasekhar limit of $\sim 1.4\
M_{\odot}$, prior to explosion? In the so-called ``single-degenerate''
scenario (Nomoto et al. 1984), the source of the accreted material is
provided by interaction with a non-degenerate companion, such as a
low-mass main sequence star, a subgiant, or a giant star (see the
reviews by Branch et al. 1995, Livio 2001). Under the
``double-degenerate'' scenario (Iben \& Tututkov 1984, Webbink 1984) the
explosion is triggered by the merger of two degenerate stars, such as
WDs or neutron stars.

The nature of SN Ia progenitors has become even more important in the
last decade because of their fundamental importance for cosmology (e.g.,
Riess et al. 1998, Perlmutter et al. 1999, Perlmutter et al. 2003). 
Unfortunately, we are in the embarrassing situation that, even as
confidence in the astounding new cosmological results increases, we
still have little knowledge of the physical origin of the luminous SNe
Ia explosions on which this result is at least partly based.

Observations of SNe Ia in the radio can provide a powerful constraint on
the Chandrasekhar mass explosion mechanism in that the mass exchange
required for the single-degenerate scenario should establish an enhanced
CSM density, however tenuous, near the SN progenitor, while the
double-degenerate scenario should not, unless the two coalescing WDs are
in a common envelope (see Livio 2003).  Thus, radio detection of the
blastwave interaction with such a CSM would not only support the
single-degenerate scenario, but also provide information on its extent,
density, and structure. 

To test the single-degenerate case for a symbiotic star progenitor
system (in which a red giant donates mass to the WD via a wind), Boffi
et al. (1995) modeled any putative, fast-evolving radio light curves for
the SN Ia 1986G through analogy with the radio light curves for the SN
Ib 1983N (Sramek et al. 1984, Weiler et al. 1986).  From radio
observations of SN 1986G conducted one week before optical maximum
(i.e., early enough to adequately test the Boffi \& Branch prediction),
Eck et al. (1995) concluded, based on a lack of detected radio emission,
that this SN probably did not arise from a symbiotic star system.  Eck
et al.~stressed that further searches for prompt radio emission from
other SNe Ia were clearly necessary to test this and other models.

Numerous radio observations have been obtained for 27 SN Ia using the
Very Large Array (VLA)\footnote{The VLA telescope of the National Radio
Astronomy Observatory is operated by Associated Universities, Inc. under
a cooperative agreement with the National Science Foundation.} as part
of a SN monitoring program we have been conducting for more than two
decades. In our program we have considered only SN Ia exploded in
external galaxies and we have not included any of the historical SN Ia
that have occurred in the Milky Way.  Earlier results of SN Ia
observations were presented by Weiler et al. (1986, 1989) and Weiler \&
Sramek (1988). Unlike Type II and Type Ib/c SNe, no SN Ia has yet been
detected as a radio emitter, even when observed quite promptly after
explosion (e.g., SNe 1981B and 1980N; Weiler et al. 1986, 1989) or quite
nearby (e.g., SN 1972E; Cowan \& Branch 1982, 1985, Weiler et al.
1989).  Here, we present our new results with improved sensitivity and
time and frequency sampling and discuss the possible implications for
the nature of the SN Ia progenitors.

\section{Radio Observations\label{observ}}

The sample of observed SNe Ia consists of those generally with $m \leq
14$ mag occurring between 1982 and 2002.  The sample includes the
``Branch Normal'' SNe Ia 1992A (e.g., Kirshner et al.~1993) and 1994D
(e.g., Richmond et al.~1995), as well as the overly luminous SN 1991T
(e.g., Filippenko et al.~1992a; Schmidt et al.~1994), the subluminous SN
1991bg (e.g., Filippenko et al.~1992b; Leibundgut et al.~1993), and the
peculiar SN 1986G (e.g., Phillips et al.~1987).  Two SNe, 2002bo (Szabo
et al.~2003) and 2002cv (Di Paola et al.~2002), both occurred within a
few months in the same host galaxy, NGC 3190.  

The SNe were observed with the VLA in a number of array configurations
at, primarily, 6 cm wavelength (4.8 GHz),  although observations for
several objects were also made at 20 cm (1.4 GHz), 3.6 cm (8.4 GHz), 2
cm  (15 GHz), 1.3 cm (22 GHz), and 0.7 cm (43 GHz). The techniques of
observation, editing, calibration, and error estimation are described in
previous publications on the radio emission from SNe (e.g., Weiler et
al. 1986, 1990, 1991).   For the SN 1986G in NGC 5128 observations, the
presence of the very radio bright galaxy nucleus Centarus A required the
additional analysis steps of first self-calibrating on, and then
removing from the {\it uv}-data, the components of the bright radio
galactic nucleus, before producing the final maps of the SN field.

The SNe Ia which were observed are listed in table \ref{tab1}.   These
include observations already described in Weiler et al.~(1986) and
Weiler et al.~(1989).  Column 1 lists the SN name, columns 2 and 3 list
the date and magnitude at optical maximum (if available; if not, the
date and magnitude at discovery are listed), and columns 4 and 5  give
the right ascension (R.~A.) and declination (Decl.) in epoch J2000
coordinates which were used for the radio observations. Columns 6 and 7
list the parent galaxy name and Hubble type.  The radio results are
listed in Table \ref{tab2}.   Column 1 of that table lists the SN name,
and column 2 the date of observation. Column 3 lists the estimated age
of the SN in days after explosion, and column 4 the VLA configuration of
the observation. Columns 5--10 list the measured rms (1$\sigma$) error,
in mJy, in the resulting maps at 20, 6, 3.6, 2, 1.3, and 0.7 cm
wavelengths, respectively. 

\begin{figure}
\epsscale{0.80}
\plotone{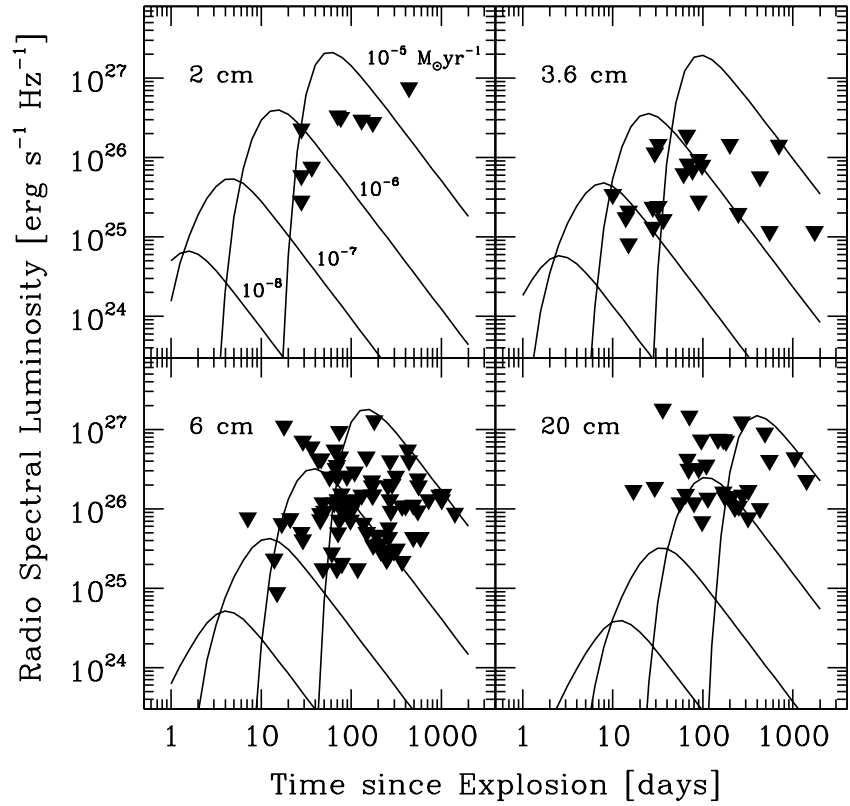}
\caption{The upper limits ($2 \sigma$) for all observed SNe Ia at 2,
3.6, 6, and 20 cm wavelength.  Shown as examples are model radio light
curves appropriate for SNe Ib/c (see text), assuming mass-loss rates
associated with the progenitor systems of $10^{-8}$, $10^{-7}$,
$10^{-6}$, and $10^{-5}$ $M_{\odot}$ yr$^{-1}$ in a stellar wind with
speed of $w_{\rm wind} = 10$ km s$^{-1}$. 
\label{fig1}
}
\end{figure}

\begin{figure}
\epsscale{0.80}
\plotone{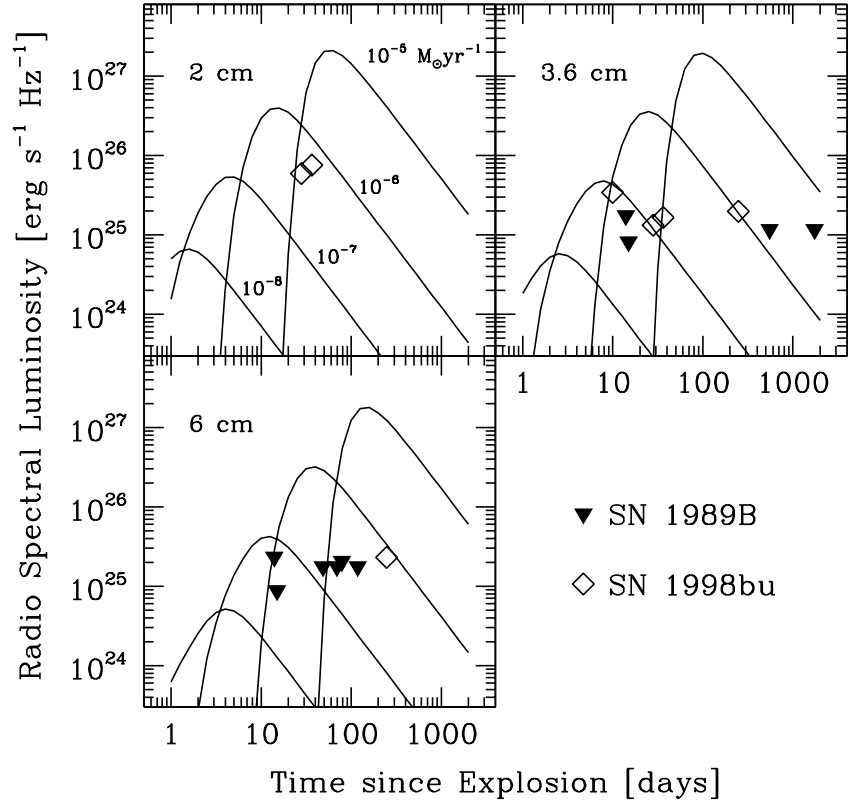}
\caption{The upper limits ($2 \sigma$) for SN~1989B at 2, 3.6, and 6 cm
wavelength ({\it filled triangles}) and SN~1998bu at 3.6 and 6 cm
wavelength ({\it open diamonds}), compared with model radio light curves
appropriate for SNe Ib/c (see text), assuming mass-loss rates associated
with the progenitor systems of $10^{-8}$, $10^{-7}$, $10^{-6}$, and
$10^{-5}$ $M_{\odot}$ yr$^{-1}$ in a stellar wind with speed $w_{\rm
wind} = 10$ km s$^{-1}$. 
\label{fig2}
}
\end{figure}

In Figure \ref{fig1} we present the $2 \sigma$ spectral luminosity upper
limits for SNe Ia at the wavelengths of 20, 6, 3.6, and 2 cm. The data
were rather sparse at 1.3 cm (1 point) and 0.7 cm (1 point), so those
two frequencies are not shown. In Figure \ref{fig2} we show the limits
for two of the SNe Ia (SN~1989B and SN~1998bu) which were particularly
well observed. None of the SNe Ia was detected as a source of radio
emission, consistent with the previous results from Weiler et al. (1986,
1989).  

\section{Radio Light Curve Modeling\label{lc}}

Following Weiler et al (2002) and Sramek \& Weiler (2003) we adopt a
parameterized model that, in view of the non-detection results, has been
simplified to include only the intrinsic synchrotron emission from
relativistic electrons created at the SN shock front and possible
free-free (f-f) absorption by thermal electrons in a surrounding
homogeneous CSM that has been ionized by the SN explosion itself.  In
this case, we can write 

\begin{equation}
\label{eq1}
S(\mbox{mJy}) = K_1 \left(\frac{\nu}{\mbox{5\ GHz}}\right)^{\alpha}
\left(\frac{t-t_0}{\mbox{1\ day}}\right)^{\beta} e^{\tau_{{\rm CSM}_{\rm external}}}  
\end{equation} 

\noindent with  

\begin{equation}
\label{eq2}
\tau_{{\rm CSM}_{\rm external}} = K_2
\left(\frac{\nu}{\mbox{5 GHz}}\right)^{-2.1}
\left(\frac{t-t_0}{\mbox{1\ day}}\right)^{\delta}
\end{equation}

\noindent with $K_1$ and $K_2$ corresponding, formally but not
necessarily physically, to the flux density ($K_1$) and homogeneous
($K_2$) absorption at 5~GHz one day after the explosion date, $t_0$. 
The term $\tau_{{\rm CSM}_{\rm external}}$ is produced by an ionized
medium that homogeneously covers the emitting source (``homogeneous
external absorption'') and is near enough to the SN progenitor that it
is altered by the rapidly expanding SN blastwave. The radial density
($\rho$) distribution of this homogeneous, external absorbing medium,
if established by a constant mass-loss rate ($\dot M$), constant
velocity ($w_{\rm wind}$) pre-SN stellar wind, is $r^{-2}$ (\ie $\rho
\propto \frac{\dot M}{w_{\rm wind}~r^2}$).  The value of $\delta$ in
Equation \ref{eq2} describes the actual radial density, if different
from  $r^{-2}$, for a constant shock velocity.  The absorbing medium is
assumed to be purely thermal, singly ionized gas, which absorbs via
free-free (f-f) transitions ! with frequency dependence $\nu^{-2.1}$ in
the radio. 

Since the f-f optical depth outside the blastwave emitting region is
proportional to the integral of the square of the CSM density over the
radius and, in the simple model (Chevalier 1982a,b) the CSM density
decreases as $r^{-2}$, the external optical depth will be proportional
to $r^{-3}$. With the blastwave radius increasing as a power of time, $r
\propto t^m$ (with $m \leq 1$; i.e., $m = 1$  for undecelerated
blastwave expansion), it follows that the deceleration parameter,  $m$,
is

\begin{equation}
\label{eq3}
m = -\delta / 3.
\end{equation}

The success of the basic parameterization and modeling has been shown in
the good correspondence between the model fits and the data for all RSN
types: e.g.,  Type Ib SN1983N (Sramek et al.~1984), Type Ic SN1990B
(VanDyk et al 1993a), Type II SN1979C  (Weiler et al 1991, 1992a, Montes
et al. 2000) and SN1980K (Weiler et al. 1992b, Montes et al. 1998), and
Type IIn SN1988Z (VanDyk et al. 1993b, Williams et al. 2002). 

For the case of a steady pre-SN stellar wind Weiler et al (1986, 2002),
and Weiler, Panagia \& Montes (2001) have shown that the mass-loss rate
can be derived directly from the measured (f-f) optical depth as: 

\begin{eqnarray}
\label{eq4}
\frac{\dot M ({\rm M_\odot} ~ {\rm yr}^{-1})}{( w_{\rm wind} / 10\ {\rm km\ s}^{-1} )} = 
3.0 \times 10^{-6}\ <\tau_{{\rm eff}}^{0.5}> \ m^{-1.5} \times \nonumber \\ {\left(\frac{v_{\rm i}}{10^{4}\ 
{\rm km\ s}^{-1}}\right)}^{1.5}{\left(\frac{t_{\rm i}}{45\ 
{\rm days}}\right) }^{1.5} {\left(\frac{t}{t_{\rm i}}\right) }^{1.5 m}{\left(\frac{T}{10^{4}\ 
{\rm K}} \right)}^{0.68}
\end{eqnarray}
\medskip

\noindent where, since the appearance of optical lines for measuring SN
ejecta velocities is often delayed a bit relative to the time of the
explosion, they arbitrarily take $t_{\rm i}$ = 45 days.  Because
observations have shown that, generally, 0.8 $\le m \le$ 1.0 and from
equation \ref{eq4} $\dot M \propto t_{\rm i}^{1.5(1-m)}$, the dependence
of the calculated mass-loss rate on the date $t_{\rm i}$ of the initial
ejecta velocity measurement is weak ($\dot M \propto t_{\rm i}^{<0.3}$),
so that the best optical or VLBI velocity measurements available can be
used without worrying about the deviation of the exact measurement epoch
from the assumed 45 days after explosion.  For convenience, and because
many SN measurements indicate velocities of $\sim10,000$ \kms, one
usually assumes $v_{\rm i} = v_{\rm blastwave} = 10,000$ \kms, CSM
temperature $T = 2 \times 10^4$ K,  
appropriate for a RSG wind),  time $t = (t_{\rm 6\ cm\ peak} - t_0)$
days from best fits to the radio data for  each RSN, and $m$ from
Equation \ref{eq3}. 

The assumed pre-SN wind velocity, $w_{\rm wind} = 10$ \kms,  appropriate
for a red supergiant (RSG) wind can also be generally applied to red
giant companions in symbiotic systems and to recurrent novae (e.g., the
CSM from the red giant wind in RS Oph is $\lesssim 20$ \kms; Hachisu \
Kato 2001). However, one should note that Shore et al. (1996) assume for
RS Oph a red giant terminal velocity of $50--100$ \kms, and Solf, B\"ohm
\& Raga (1986) find bipolar winds in symbiotic systems with speeds in
excess of 100 \kms.   For the case of possible dwarf and subdwarf winds,
one would expect velocities of the order of their escape velocities,
i.e. few hundred \kms, but also much lower mass loss rates.  Therefore,
in these cases any mass transfer induced by stellar winds would have no
effect on building  up a dense CSM environment, and can safely be
neglected in our discussion.

\section{Application to Type Ia Supernovae\label{apply}}

Since the overall shape of the optical light curves of SNe Ia are rather
similar to those of SNe Ib/c (although SNe Ib/c are generally
$\sim$1--1.5 mag fainter than SNe Ia; see e.g., Leibundgut et al.~1991),
suggesting comparable envelope masses and structures, we will adopt for
our analysis average parameters measured for SNe Ib/c (see Weiler et
al.~2002), namely $\alpha = -1.1$, $\beta = -1.5$, and $\delta =
-2.6$.   This is very similar to the approach taken by Boffi et al.
(1995), where they argued that, in particular for a symbiotic stellar
system progenitor scenario, many of the parameters expected for SNe Ia
(specifically, SN 1986G in their case) may be analogous to those for SNe
Ib/c.

As discussed earlier, the SN radio emission is a function of the CSM
density and, hence, is proportional to the ratio of the mass-loss rate
to the wind velocity, $\dot M/w$. The theory developed by Chevalier
(1982a,b) provides a functional dependence between the intrinsic (\ie
before external f-f absorption by an ionized CSM) radio luminosity and
the density of the CSM, of the form (Weiler et al. 1989, 2002)

\begin{eqnarray}
\label{eq5}
L_{\rm intrinsic}(\nu) & \propto  (\dot
M/w)^{(\gamma-7+12m)/4}m^{(5+\gamma)/2} \times \nonumber \\ & t^{-(\gamma+5-6m)/2}\nu^{-(\gamma-1)/2}
,
\end{eqnarray}

\medskip
\noindent where

\begin{equation}
\label{eq6}
\gamma = 1 - 2 \alpha.
\end{equation}

\noindent For the assumed $\alpha = -1.1$ ($\gamma = 3.2$), $\delta =
-2.6$ ($m = -\delta/3 = 0.87$), and at $\nu = 5$ GHz, this becomes

\begin{equation}
\label{eq7}
L_{\rm intrinsic}(\nu) \propto (\dot M/w)^{1.65}t^{-1.5}\nu^{-1.1}.
\end{equation}

\noindent Multiplying by the attenuation produced by
external (f-f) absorption, it becomes analogous to Equation (1), but
when expressed in absolute units is

\begin{eqnarray}
\label{eq8}
{\frac{L}{10^{26}~{\rm erg\ s^{-1}\ Hz^{-1}}}}  =  \Lambda\ 
{\left(\frac{\dot M ({\rm M_\odot} ~ {\rm yr}^{-1})}{ w_{\rm wind} / 10\ {\rm
km\ s}^{-1} }\right)}^{1.65}  \times \nonumber \\  {\left(\frac{{\nu}}{{\rm 5 GHz}}\right)}^{-1.1} {\left(\frac{{t-t_0}}{{\rm days}}\right)}^{-1.5} e^{-\tau_{{\rm CSM}_{\rm external}}}
\end{eqnarray}

\medskip
The parameter $\Lambda$ is a proportionality constant which is not
provided by theory and must be calibrated empirically from radio
observations of SNe Ib/c.  Since we are dealing with sources that are
intrinsically faint, we  calibrate $\Lambda$ using the best measured
faint SN Ib/c, namely SN 1983N (Sramek et al. 1984, Weiler et al. 1989,
2002), obtaining

\begin{equation}
\label{eq9}
\Lambda = 1285 \pm 245.
\end{equation}

The quoted error  corresponds to the combination in quadrature of the 
uncertainty of the light curve fit ($\sim14\%$) as given by Weiler et
al. (1986) and an estimated uncertainty in the distance to M83 of
~$\sim7\%$ (Thim et al. 2003).  Also, one should be aware that
appreciably different values of $\Lambda$ would be obtained from using
different SNe Ib/c listed in table 3 of Weiler et al. (2002) for its
estimation, with the bright SNe providing systematically lower values
than the fainter ones, with a range of a factor of 10 between the
extremes.   However, because of the functional dependence of $L$ on
$\dot M/w_{\rm wind}$, such a spread would result in an overall
uncertainty in $\dot M$ of at most a factor of two.

For each SN the upper limit to the mass-loss rates were estimated by
direct comparisons of the radio luminosity upper limits with a set of
theoretical light curves calculated for all relevant epochs and for
values of the parameter $\dot M/w_{\rm wind}$ between $10^{-9}$ and
$10^{-6}$ \msun\ yr$^{-1}$ km$^{-1}$ s at logarithmic steps of 0.05.  In
the case of SNe with observations at different frequencies, the overall
upper limit to $\dot M/w_{\rm wind}$ was taken as the minimum value
among those determined for each frequency.  Finally, the mass-loss rates
were calculated assuming the wind velocity to be $w=10$ \kms, as
appropriate for winds from red giants and RSGs, as well as from binary
systems with total mass of a few solar masses and separations of a few
AU.

Table \ref{tab3} lists upper limits to the mass-loss rates from the SN
Ia progenitor systems. Column 1 lists the SN name, and column 2 its
distance taken from direct Cepheid determinations, whenever possible, or
from the Revised Shapley-Ames Catalog of Bright Galaxies (Sandage \&
Tammann 1987) rescaled to a value of the Hubble constant H$_0 = 63$
\kms\ Mpc$^{-1}$ (Panagia 2003). Columns 3, 4, and 5 list the epoch, the
wavelength, and the $2\sigma$ limit  to the spectral luminosity that 
yielded the lowest mass-loss rate limit, and the last column (6) lists
the $2\sigma$ limit to the mass-loss rate, $\dot M$.

\section{Discussion\label{discuss}}

Examination of Table \ref{tab3} shows that our upper limits are
generally consistent with pre-SN mass-loss rates lower than $10^{-6}$
M$_\odot$ yr$^{-1}$ associated with the progenitor system, with almost
half (actually 12 out of 27) of the  $2 \sigma$ limits falling below
$\sim 4\times 10^{-7}$ M$_\odot$ yr$^{-1}$. In particular, the most
stringent upper limits are provided by the observations of SNe 1989B and
1998bu, as shown in Figure \ref{fig2}.   Not surprisingly, it also
appears that most of the lowest upper limits are found for relatively
nearby SNe Ia observed at early epochs, \eg $\sim10$--50 days after
explosion or about $-10$ to $+30$ days relative to the optical maximum,
and at frequencies 5--8.3 GHz ($\lambda = 3.6$--6 cm).  This is because
the VLA sensitivity is highest in the 5 and 8.3 GHz bands and the
expected radio light curves at these frequencies peak around epochs of
10 days for mass-loss rates of $\sim 10^{-7}$ \msun\ yr$^{-1}$.  On this
basis, we argue that future searches should focus their efforts in the 5
and 8.3 GHz bands, making prompt observations of SNe Ia soon after they
are discovered, and repeating the  observations about every ten days
until one or two months past the optical maximum.  Additional
observations at later times and at longer wavelengths would also be
useful to check on the possibility that a non-detection may be due to
strong (f-f) absorption (\ie indicating a {\it high} mass-loss rate),
rather than merely to intrinsically faint emission.

We can derive a more stringent limit if we assume that  {\it all} SNe Ia
have the  same type of progenitors and that they reach the explosion
along  identical evolutionary paths.  In this case, we can combine the
upper limits derived for individual objects, because they could be
considered as the ``errors'' in a series of independent measurements,
\ie $\sigma{^2_{\rm combined}}=1/\Sigma_i (1/\sigma{^2_i})$.  In this
way, we obtain a combined $2\sigma$ upper limit of $\lesssim
2.6\times10^{-8}$ M$_\odot$ yr$^{-1}$.

Such low mass-loss rates as we have estimated generally rule out any red
giant or red supergiant SN progenitors (consistent with current
theories), and also exclude any SN Ia progenitor models that invoke a
stellar wind from a (massive) companion to provide the required
accretion rate onto a WD.  Nomoto et al. (1984) calculated that only
accretion rates appreciably greater than $\sim4 \times 10^{-8}$
M$_\odot$ yr$^{-1}$ can lead to a mass increase of the accreting WD
resulting in a SN Ia explosion. More recently, Prialnik \& Kovetz (1995)
estimate that accretion rates $\gtrsim 10^{-7}$ M$_\odot$ yr$^{-1}$ are
needed for a WD to increase its mass so as to exceed the Chandrasekhar
mass.

Since accretion from a companion star's wind is a rather inefficient
process, with $\sim$10\% of the wind material being accreted onto a WD
Yungelson et al. (1995), only stellar companions with mass-loss rates
$\gtrsim 4-10\times10^{-7}$ M$_\odot$ yr$^{-1}$ would be able to satisfy
the requirement.  These $\dot{M}$ values are much greater ($\sim$10--25
times) than the best individual, and particularly the aggregate, upper
limits derived from our SN Ia radio observations. Thus, we support the
conclusion reached by Boffi et al. (1995) and Eck et al. (1995) that
symbiotic systems  are unlikely to be SN progenitors.

An alternative to wind accretion onto the WD is Roche lobe overflow from
a giant, subgiant, or main sequence companion (see Branch et al. 1995,
Livio 2001), so that the process is more gradual and efficient.  In this
case our best combined upper limit, \ie  $2.6\times10^{-8}$ M$_\odot$
yr$^{-1}$, implies that such a mass transfer must have an efficiency of
$\gtrsim 60$--80~\% (depending on the adopted limiting accretion rate
for WDs to become SNe Ia) to avoid leaving a residual CSM from which
synchrotron emission would be detectable at current radio limits.

Double degenerate models for SN Ia events, in which the explosion is
triggered by the merger of two WDs in a binary system are, of course,
still consistent with our observations.

One should keep in mind that our mass-loss rate estimates are dependent
on the assumptions that SNe Ia radio light curves will be very similar
to those of SNe Ib/c and that $w_{\rm wind} = 10$ \kms, $v_{\rm i} =
10^4$ \kms, and $T = 2 \times 10^4$ K, with dependences shown in
Equations \ref{eq4} and \ref{eq8}. Even if higher values for any of
these quantities are more appropriate or realistic, and will therefore
yield higher mass-loss limits and correspondingly less stringent limits
on the properties of the progenitor systems, we can definitely rule out
the symbiotic system scenario based on our complete dataset.

\section{Conclusions\label{conclude}}

No radio emission has been found from 27 SNe Ia observed over two
decades with the VLA. It is clear that the CSM environment is far more
tenuous than that of SNe II, or even SNe Ib/c.  Using model predictions
of radio emission from SNe and assuming that the radio properties of SNe
Ia would be relatively similar to those of SNe Ib/c (at least for some
progenitor system models), but with a wind-established CSM from a less
massive pre-SN system, we can place constraints on the mass-loss rate
from such systems. If SNe Ia originate from mass accretion from a
massive companion's wind onto a WD, we can establish a stringent limit
of $\dot M \lesssim 3 \times 10^{-8}$ M$_{\odot}$ yr$^{-1}$ in the best
cases.  This severely limits the possibility that the progenitors are
symbiotic systems, where the companion is a red giant or supergiant
star. However, high-efficiency  mass transfer through Roche lobe
overflow from a lower mass companion or the double degenerate scenario,
involving the merger of a two WDs or a WD and a neutron star, cannot be
ruled out. 

We note that even in the most unusual case so far, of the recent SN Ia
2002ic, which exhibited optical evidence for a substantial CSM (Hamuy et
al. 2003), it was  concluded that the double degenerate scenario was the
most consistent model for the progenitor of that event (Hamuy et al.
2003, Livio \& Riess 2003).  Attempts to detect radio emission from SN
2002ic with the VLA were unsuccessful, likely due to the SN's large
distance (Stockdale et al. 2003, Berger et al. 2003). Another recent
example appears to be SN 2005gj (Prieto et al.~2005), which has also not
been detected using the VLA (Soderberg \& Frail 2005).  We further note,
of course, that such events are likely intrinsically quite rare in
general.

Clearly, additional observations of new SNe Ia should be made with the
VLA as early and as deeply after discovery as possible, to help further
constrain the nature of SN Ia progenitors.  We continue to attempt
observations such as these.  However, with the current VLA, based on the
results presented in  this paper, realistically only the next Local
Group SN Ia, or possibly a SN in a nearby galaxy group, will provide
greatly improved upper limits. The plans for the Expanded 
VLA\footnote{http://www.nrao.edu/evla/.}, and the far greater
sensitivity that the upgrade will produce, will likely soon afford us
with  an unprecedented opportunity to detect radio emission from SN Ia
and help define their progenitor systems.  Considering the cosmological
implications, such future observations are essential.

\acknowledgments

KWW thanks the Office of Naval Research (ONR) for the 6.1 funding
supporting this research.  CJS is a Cottrell Scholar of Research
Corporation and work on this project has been supported by the NASA
Wisconsin Space Grant Consortium. Additional information and data on RSNe
can be found at \\{\it http://rsd-www.nrl.navy.mil/7210/7213/index.htm}
and linked pages.


\begin{deluxetable}{lcccccc}
\tablenum{1}
\tablewidth{6.9in}
\tablecolumns{7}
\tablecaption{Observed SNe}
\tablehead{
\colhead{SN} & \colhead{Date of} & \colhead{Magnitude}
& \multicolumn{2}{c}{SN Position (J2000.0)} & 
\multicolumn{2}{c}{Parent Galaxy} \\
\colhead{Name} & \colhead{Optical} & \colhead{At} & 
\multicolumn{2}{c}{\hrulefill} & \multicolumn{2}{c}{\hrulefill} \\
\colhead{} & \colhead{Maximum\tablenotemark{a}} & \colhead{Maximum\tablenotemark{a}} & 
\colhead{R.~A.} & \colhead{Decl.} & \colhead{Name} & \colhead{Type}}
\startdata
1980N & 1980 Dec 11   & 12.5B      & $03^h 22^m 59{\fs}8\ $ $\pm1{\fs}3$  
& $-37\arcdeg 12\arcmin 48\arcsec $ $\pm 15\farcs0$ & NGC 1316 & S0\\
1981B & 1981 Mar 12   & 12.0B	  & $12\ \ 34\ \ 29.57\ $  $\pm 0.07$  &
$ +02\ 11\ 59.3\ $ $\pm 1.0$ & NGC 4536  & Sbc \\
1982E & $\le$1982 Mar & $\le$14pg & $03\ \ 26\ \ 40.41\ $  $\pm 0.54$ 
& $-21\ 17\ 13.8\ $ $\pm 8.0$ & NGC 1332 & S0 \\
1983G & 1983 Apr 09   & 12.9B     & $12\ \ 52\ \ 21.0\ $ $\pm 1.0$  &
$ -01\ 12\ 12\ $ $\pm 15.0$  & NGC 4753 & S0 pec\\
1984A & 1984 Jan 16   & 12.4B     & $12\ \ 26\ \ 55.73\ $ $\pm 0.06$  &
$ +15\ 03\ 17.1\ $ $\pm 1.0$  & NGC 4419 & SBab\\
1985A & $\le$1985 Jan & $\le$14.5pg & $09\ \ 13\ \ 42.43$ $\pm 0.30$ & 
$+76\ 28\ 23.8$ $\pm 1.0$ & NGC 2748 & Sc \\
1985B & $\le$1985 Jan & $\le$13.0V & $12\ \ 02\ \ 43.94$ $\pm 0.07$ & 
$+01\ 58\ 45.3$ $\pm 1.0$ & NGC 4045 & Sbc \\
1986A & 1986 Feb 07 & 14.4B & 10 \ 46 \ 36.59 $\pm$ 0.07      & 
+13 \ 45 \ 00.7  $\pm$ 1.0   & NGC 3367 & SBc \\
1986G & 1986 May 11 & 12.5B & 13 \ 25 \ 36.51 $\pm$ 0.07      & 
$-$43 \ 01 \ 54.3  $\pm$ 1.0   & NGC 5128 & S0+Spec \\
1986O & $\sim$1986 Dec 20 & $\sim$14.0V & 06 \ 25 \ 58.0 $\pm$ 0.58      &
$-$22 \ 00 \ 42  $\pm$ 8.0   & NGC 2227 & SBcd \\
1987D & 1987 Apr 17 & 13.7B & 12 \ 19 \ 41.10 $\pm$ 0.07      & 
+02 \ 04 \ 26.6  $\pm$ 1.0   & M+00$-$32$-$01 & Sbc \\
1987N & $\le$1987 Dec & $\le$13.4V & 23 \ 19 \ 03.42 $\pm$ 0.07      & 
$-$08 \ 28 \ 37.5  $\pm$ 1.0   & NGC 7606 & Sb \\
1989B & 1989 Feb 06 & 12.5B & 11 \ 20 \ 13.93 $\pm$ 0.07      & 
+13 \ 00 \ 19.3  $\pm$ 1.0   & NGC 3627 & Sb \\
1989M & $\sim$1989 Jun & $\le$12.1B & 12 \ 37 \ 40.75 $\pm$ 0.07      & 
+11 \ 49 \ 26.1  $\pm$ 1.0   & NGC 4579 & Sab \\
1990M & $\sim$1990 Jun & $\le$13.4V & 14 \ 08 \ 29.3\ $\pm$ 0.1 \      & 
$-$05 \ 02 \ 36 \ \  $\pm$ 1.0   & NGC 5493 & S0 \\
1991T & 1991 Apr 28.5 & 11.64B & 12 \ 34 \ 10.20 $\pm$ 0.07      & 
+02 \ 39 \ 56.4  $\pm$ 1.0   & NGC 4527 & Sb \\
1991bg & 1991 Dec 14.7 & 13.95V & 12 \ 25 \ 03.70 $\pm$ 0.07      & 
+12 \ 52 \ 15.6  $\pm$ 1.0   & NGC 4374 & E1 \\
1992A & 1992 Jan 16 & 12.78V & 03 \ 36 \ 27.41 $\pm$ 0.07      & 
$-$34 \ 57 \ 31.4  $\pm$ 1.0   & NGC 1380 & S0/Sa \\
1994D & 1994 Mar 22 & 11.85V & 12 \ 34 \ 02.40 $\pm$ 0.007 & 
+07 \ 42 \ 05.7 $\pm$ 0.1 & NGC 4526 & SAB(s) \\
1995al & 1995 Nov 09 & 13.25V & 09 \ 50 \ 55.97 $\pm$ 0.06 & 
+33 \ 33 \ 09.4 $\pm$ 1.0 & NGC 3021 & SAbc: \\
1996X & 1996 Apr 18 & 13.24B & 13 \ 18 \ 01.13 $\pm$ 0.06 &
$-$26 \ 50 \ 45.3 $\pm$ 1.0 & NGC 5061 & E0 \\
1998bu & 1998 May 21 & 11.93V & 10 \ 46 \ 46.03 $\pm$ 0.03 & 
+11 \ 50 \ 07.1 $\pm$ 0.5 & NGC 3368 & SABab \\
1999by & 1999 May 10 & 13.8B & 09 \ 21 \ 52.07 $\pm$ 0.04 & 
+51 \ 00 06.6 $\pm$ 1.0 & NGC 2841 & SA(r)b \\
2002bo & 2002 Mar 23 & 14.04B & 10 \ 18 \ 06.51 $\pm$ 0.03 & 
+21 \ 49 \ 41.7 $\pm$ 0.5 & NGC 3190 & SA(s)a \\
2002cv & 2002 May 20 & 14.8J & 10 \ 18 \ 03.68 $\pm$ 0.03 & 
+21 \ 50 \ 06.0 $\pm$ 0.5 & \nodata & \nodata \\
2003hv & 2003 Sep 08 & $<$12.5R & 03\ 04\ 09.32 $\pm$ 0.03 &
$-$26\ 05\ 07.5 $\pm$ 0.5 & NGC 1201 & SA(r)0 \\
2003if  & 2003 Sep 01 & $<$17.6R & 03\ 19\ 52.61 $\pm$ 0.03 &
$-$26\ 03\ 50.5 $\pm$ 0.5 & NGC 1302 & SAB(r)a \\
\enddata
\tablenotetext{a}{Date and magnitude at discovery if the information is
not available for the SN maximum.}
\label{tab1}
\end{deluxetable}


\begin{deluxetable}{llcccccccc}
\tablenum{2}
\tablewidth{6.2in}
\tablecaption{Observations}
\tablehead{
\colhead{} & \colhead{} & \colhead{} & \colhead{} & \multicolumn{6}{c}{Map rms (1$\sigma$)} \\
\colhead{} & \colhead{} & \colhead{} &\colhead{} & \multicolumn{6}{c}{\hrulefill} \\
\colhead{SN} & \colhead{Observation} & \colhead{Age\tablenotemark{a}} & \colhead{VLA} 
& $\sigma_{20}$ & $\sigma_6$ & $\sigma_{3.6}$ & $\sigma_2$ 
& $\sigma_{1.3}$ & $\sigma_{0.7}$ \\
\colhead{Name} & \colhead{Date} & \colhead{(days)} & \colhead{Config.} & 
\colhead{(mJy)} & \colhead{(mJy)} & \colhead{(mJy)} & \colhead{(mJy)}
& \colhead{(mJy)} & \colhead{(mJy)}
}
\startdata
 1980N  & 1981 Feb 03 & 72  & A      &\nodata & 0.20 & \nodata & \nodata & \nodata & \nodata \\
 1981B  & 1981 Mar 11 & 18  & A       &\nodata & 0.10 & \nodata & \nodata & \nodata & \nodata \\
        & 1981 Apr 09 & 46  & A       &\nodata & 0.13 & \nodata & \nodata & \nodata & \nodata \\
        & 1981 May 14 & 82  & B       &\nodata & 0.17 & \nodata & \nodata & \nodata & \nodata \\
        & 1981 Jun 19 & 117 & B       &\nodata & 0.20 & \nodata & \nodata & \nodata & \nodata \\
        & 1981 Aug 13 & 172 & B       &\nodata & 0.30 & \nodata & \nodata & \nodata & \nodata \\
        & 1981 Nov 11 & 261 & C       &\nodata & 0.07 & \nodata & \nodata & \nodata & \nodata \\
        & 1982 Feb 27 & 369 & A       &\nodata & 0.03 & \nodata & \nodata & \nodata & \nodata \\
        & 1982 Jun 25 & 489 & A       &\nodata & 0.07 & \nodata & \nodata & \nodata & \nodata \\
        & 1982 Oct 02 & 587 & A       &\nodata & 0.07 & \nodata & \nodata & \nodata & \nodata \\
        & 1983 Feb 16 & 723 & C       &\nodata & 0.20 & \nodata & \nodata & \nodata & \nodata \\
 1982E  & 1985 Dec 29 & 1417& D       & 0.18   & 0.07 & \nodata & \nodata & \nodata & \nodata \\
 1983G  & 1983 May 27 & 72  & C     &\nodata & 0.07 & \nodata & \nodata & \nodata & \nodata \\
 1984A  & 1984 Mar 05 & 430  & BnC       &\nodata & 0.10 & \nodata & \nodata & \nodata & \nodata \\
 1985A  & 1985 Feb 01 & 49& A         &\nodata & 0.07 & \nodata & \nodata & \nodata & \nodata \\
        & 1985 Feb 08 & 56& A         &0.07    &\nodata & \nodata & \nodata  & \nodata & \nodata \\
        & 1985 Feb 17 & 65& A         &0.09 & 0.07 & \nodata & \nodata & \nodata & \nodata \\
        & 1985 Mar 02 & 81& A         &0.07 & \nodata & \nodata & \nodata  & \nodata & \nodata \\
        & 1985 Apr 05 & 114& A/B      &0.08 & \nodata & \nodata & \nodata  & \nodata & \nodata \\
        & 1985 Oct 28 & 320& C/D      &0.10 & \nodata & \nodata & \nodata  & \nodata & \nodata \\
        & 1986 Jun 15 & 550& A/B      &0.24 & 0.14 & \nodata & \nodata  & \nodata & \nodata \\
        & 1987 Oct 23 & 1045& A/B      & 0.26 & 0.09 & \nodata & \nodata  & \nodata & \nodata \\
 1985B & 1985 Feb 22 & 70& A        & 0.17 & 0.19 & \nodata & \nodata & \nodata & \nodata \\
       & 1985 Mar 18 & 94& A/B      & 0.40 & 0.06 & \nodata & \nodata & \nodata & \nodata \\
       & 1985 Sep 15 & 275& C        & 0.08 & 0.05 & \nodata & \nodata & \nodata & \nodata \\
       & 1986 Apr 30 & 502& A        & \nodata & 0.16 & \nodata & \nodata & \nodata & \nodata \\
       & 1987 Sep 18 & 1008& A        & \nodata & 0.07 & \nodata & \nodata & \nodata & \nodata \\
 1986A & 1986 Feb 07 & 18& D &    \nodata & 0.22 & \nodata & \nodata & \nodata & \nodata \\
       & 1986 Feb 25 & 36& A        & 0.36 & 0.12 & \nodata & \nodata & \nodata & \nodata \\
       & 1986 Mar 16 & 55& A        & \nodata & 0.05 & \nodata & \nodata & \nodata & \nodata \\
       & 1986 Apr 03 & 73& A        & \nodata & 0.09 & \nodata & \nodata & \nodata & \nodata \\
       & 1986 Jun 15 & 146& A/B      & 0.15 & 0.09 & \nodata & \nodata & \nodata & \nodata \\
       & 1986 Oct 16 & 269& B/C      & 0.25 & 0.08 & \nodata & \nodata & \nodata & \nodata \\
       & 1987 Apr 01 & 436& D        & \nodata & 0.08 & \nodata & 0.15 & \nodata & \nodata \\
 1986G & 1986 May 14 & 21& A        & \nodata & 1.06 & \nodata & \nodata & \nodata & \nodata \\
       & 1986 May 21 & 28& A        & \nodata & 0.70 & \nodata & 3.21 & \nodata & \nodata \\
       & 1986 Jun 08 & 46& A        & 7.48 & 1.02 & \nodata & \nodata & \nodata & \nodata \\
       & 1986 Jul 06 & 74& A/B      & \nodata & 1.52 & \nodata & 1.32 & \nodata & \nodata  \\
       & 1986 Sep 20 & 150& C        & \nodata & 2.49 & \nodata & 1.27 & \nodata & \nodata \\
       & 1987 Jan 04 & 256& C        & \nodata & 2.71 & \nodata & 1.38 & \nodata & \nodata \\
       & 1987 Oct 23 & 548& A/B      & 5.06 & 1.31 & \nodata & 4.47 & \nodata & \nodata  \\
       & 1989 Apr 06 & 1079& B        & \nodata & 4.08 & \nodata & \nodata & \nodata & \nodata \\
 1986O & 1987 Feb 12 & 72& C/D      & 0.80 & 0.07 & \nodata & 0.18 & \nodata & \nodata \\
       & 1987 Apr 11 & 130& D        & \nodata & 0.08 & \nodata & 0.16 & \nodata & \nodata \\
       & 1987 May 24 & 173& D        & \nodata & 0.08 & \nodata & 0.15 & \nodata & \nodata \\
       & 1987 Aug 28 & 269& A        & \nodata & 0.07 & \nodata & \nodata & \nodata & \nodata \\
       & 1988 Apr 03 & 488& C        & 0.49 & 0.06 & \nodata & \nodata & \nodata & \nodata \\
       & 1989 Jul 17 & 958& C        & \nodata & 0.08 & \nodata & \nodata & \nodata & \nodata \\
 1987D & 1987 May 15 & 46& D        & \nodata & 0.20 & \nodata & \nodata & \nodata & \nodata \\
       & 1987 Jun 04 & 66& D        & \nodata & 0.26 & \nodata & \nodata & \nodata & \nodata \\
       & 1987 Jun 21 & 83& A        & \nodata & 0.06 & \nodata & \nodata & \nodata & \nodata \\
       & 1987 Sep 18 & 172& A        & \nodata & 0.07 & \nodata & \nodata & \nodata & \nodata \\
       & 1988 May 29 & 426& C/D      & \nodata & 0.26 & \nodata & \nodata & \nodata & \nodata \\
 1987N & 1987 Dec 20 & 37& B        & 0.13 & 0.10 & \nodata & \nodata & \nodata & \nodata \\
       & 1988 Jan 12 & 60& B        & 0.10 & 0.08 & \nodata & \nodata & \nodata & \nodata \\
       & 1988 Feb 01 & 76& B        & 0.11 & 0.09 & \nodata & \nodata & \nodata & \nodata \\
       & 1988 Mar 31 & 135& C        & 0.23 & 0.07 & \nodata & \nodata & \nodata & \nodata \\
       & 1988 Apr 11 & 146& C        & 0.22 & 0.40 & \nodata & \nodata & \nodata & \nodata \\
       & 1988 Aug 22 & 279& D        & \nodata & 0.08 & \nodata & \nodata & \nodata & \nodata \\
       & 1989 Apr 24 & 524& B        & \nodata & 0.06 & \nodata & \nodata & \nodata & \nodata \\
 1989B & 1989 Feb 02\tablenotemark{b} & 10& A & \nodata & 0.08 & 0.06 & \nodata  & \nodata & \nodata \\ 
       & 1989 Feb 03 & 11& A          & \nodata & 0.03 & 0.03 & \nodata & \nodata & \nodata \\ 
       & 1989 Mar 06 & 42& A/B      & \nodata & 0.06 & \nodata & \nodata & \nodata & \nodata \\
       & 1989 Mar 27 & 63& B        & \nodata & 0.06 & \nodata & \nodata & \nodata & \nodata \\
       & 1989 Apr 06 & 73& B        & \nodata & 0.07 & \nodata & \nodata & \nodata & \nodata \\
       & 1989 May 15 & 112& B/C      & \nodata & 0.06 & \nodata & \nodata & \nodata & \nodata \\
       & 1990 Jul 22 & 545& B        & \nodata & \nodata & 0.04 & \nodata & \nodata & \nodata \\
       & 1993 Oct 25 & 1736& C/D      & \nodata & \nodata & 0.04 & \nodata & \nodata & \nodata \\
       & 2003 May 26 & 5236& A        & \nodata & \nodata & \nodata & \nodata & \nodata & \nodata \\
 1989M & 1989 Jul 17 & 60& C        & \nodata & 0.13 & \nodata & \nodata & \nodata & \nodata \\
       & 1989 Sep 04 & 109& C        & \nodata & 0.10 & \nodata & \nodata & \nodata & \nodata \\
       & 1989 Oct 24 & 159& C/D      & \nodata & 0.07 & \nodata & \nodata & \nodata & \nodata \\
       & 1989 Dec 21 & 217& D        & \nodata & 0.05 & \nodata & \nodata & \nodata & \nodata \\
       & 1990 Feb 13 & 271& D/A      & \nodata & 0.08 & \nodata & \nodata & \nodata & \nodata \\
       & 1990 May 29 & 376& A        & \nodata & 0.15 & \nodata & \nodata & \nodata & \nodata \\
\enddata
\end{deluxetable}
\begin{deluxetable}{llcccccccc}
\tablenum{2}
\tablewidth{6.2in}
\tablecaption{Observations (Cont.)}
\tablehead{
\colhead{} & \colhead{} & \colhead{} & \colhead{} & \multicolumn{6}{c}{Map rms (1$\sigma$)} \\
\colhead{} & \colhead{} & \colhead{} &\colhead{} & \multicolumn{6}{c}{\hrulefill} \\
\colhead{SN} & \colhead{Observation} & \colhead{Age\tablenotemark{a}} & \colhead{VLA} 
& $\sigma_{20}$ & $\sigma_6$ & $\sigma_{3.6}$ & $\sigma_2$ 
& $\sigma_{1.3}$ & $\sigma_{0.7}$ \\
\colhead{Name} & \colhead{Date} & \colhead{(days)} & \colhead{Config.} & 
\colhead{(mJy)} & \colhead{(mJy)} & \colhead{(mJy)} & \colhead{(mJy)}
& \colhead{(mJy)} & \colhead{(mJy)}
}
\startdata
 1990M & 1990 Jun 29 & 42& A/B      & \nodata & \nodata & 0.04 & \nodata & \nodata & \nodata \\
       & 1990 Dec 14 & 210& C        & \nodata & \nodata & 0.04 & \nodata & \nodata & \nodata \\
 1991T & 1991 May 08 & 24& D        & \nodata & \nodata & 0.05 & 0.06 & \nodata & \nodata \\
       & 1991 Jul 09 & 86& A        & \nodata & \nodata & 0.06 & 0.20 & \nodata & \nodata \\
       & 1993 Feb 02 & 660& A/B      & 0.05 & \nodata & 0.05 & \nodata & \nodata & \nodata \\
 1991bg & 1991 Dec 26& 26& B       & 0.26 & \nodata & 0.16 & \nodata & \nodata & \nodata \\
       & 1993 Feb 02 & 430& A/B      & 0.14\tablenotemark{c} & \nodata & 0.08 & \nodata & \nodata & \nodata \\
       & 1993 Oct 25 & 695& C/D      & \nodata & \nodata & 0.20\tablenotemark{b} & \nodata & \nodata & \nodata \\
 1992A & 1992 Jan 27 & 25& B/C      & \nodata & 0.03 & \nodata & \nodata & \nodata & \nodata \\
       & 1992 Oct 09 & 281& A        & \nodata & 0.15 & \nodata & \nodata & \nodata & \nodata \\
       & 1993 Feb 05 & 400& A/B      & \nodata & 0.08 & \nodata & \nodata & \nodata & \nodata \\
 1994D & 1994 May 04 & 57& A        & \nodata & 0.06 & \nodata & \nodata & \nodata & \nodata \\
 1995al & 1995 Nov 08 & 13& B       & 0.08 & \nodata & \nodata & \nodata & \nodata & \nodata \\
 1996X & 1996 May 29 & 55& DnC        & \nodata & \nodata & 0.09 & \nodata & \nodata & \nodata \\
 1998bu & 1998 May 13 & 6& A       & \nodata & \nodata & 0.07 & \nodata  & \nodata & \nodata \\
        & 1998 May 31 & 24& A       & \nodata & \nodata & 0.04 & 0.18 & \nodata & \nodata \\
        & 1998 Jun 09 & 33& AnB       & \nodata & \nodata & 0.05 & 0.23 & \nodata & \nodata \\
        & 1999 Jan 07 & 245& C       & \nodata & 0.07    & 0.06 & \nodata & \nodata & \nodata \\
 1999by & 1999 May 07 & 11& D       & \nodata & \nodata & 0.07 & \nodata & \nodata & \nodata \\
        & 1999 May 24 & 28& D       & \nodata & \nodata & 0.08 & \nodata & \nodata & \nodata \\
 2002bo & 2002 May 21 & 73& AnB       & \nodata & \nodata & 0.06 & 0.28 & 0.36 & 0.62 \\
           & 2002 Jun 12 & 95& B       & 0.06 & 0.08 & 0.07 & \nodata & \nodata & \nodata \\
 2002cv & 2002 May 21 & 19 & AnB & \nodata & \nodata & 0.06 & 0.28 & 0.36 & 0.62 \\
           & 2002 Jun 12 & 41& B       & 0.06 & 0.08 & 0.07 & \nodata & \nodata & \nodata \\
 2003hv & 2003 Oct 21 & 57& B & \nodata & \nodata & 0.05 & \nodata & \nodata & \nodata \\
 2003if   & 2003 Oct 21 & 64& B & \nodata & \nodata & 0.05 & \nodata & \nodata & \nodata \\
\enddata
\tablenotetext{a}{The explosion date is taken to be 18 days before the
date of the optical maximum (Goldhaber et al. 2001).}
\tablenotetext{b}{Observations were graciously contributed by R.~Brown.}
\tablenotetext{c}{Severely confused by the southern radio lobe from the
host galaxy, NGC 4374.}
\label{tab2}
\end{deluxetable}

\begin{deluxetable}{lccccc}
\tablenum{3}
\tablewidth{5.0in}
\tablecaption{Lowest Upper Limits to SN Ia Progenitor Mass-Loss Rates}
\tablehead{
\colhead{SN} & \colhead{Distance} & \colhead{Epoch} &  \colhead{Wavelength} & \colhead{Radio Luminosity\tablenotemark{a}} & \colhead{$\dot M$ \tablenotemark{b}} \\
\colhead{} & \colhead{(Mpc)} & \colhead{(days)} &  \colhead{(cm)} &\colhead{(erg s$^{-1}$ Hz$^{-1}$)} 
& \colhead{($M_{\odot}$ yr$^{-1}$)}
}
\startdata
1980N & 23.3  & 71   & 6   & $2.5 \times 10^{26}$ & $1.1 \times 10^{-6}$ \\
1981B & 16.6  & 17   & 6   & $6.5 \times 10^{25}$ & $1.3 \times 10^{-7}$ \\
1982E & 23.1  & 1416 & 20  & $2.3 \times 10^{26}$ & $7.3 \times 10^{-6}$ \\
1983G & 17.8  & 71   & 6   & $5.0 \times 10^{25}$ & $4.1 \times 10^{-7}$ \\
1984A & 17.4  & 74   & 6   & $7.1 \times 10^{25}$ & $5.3 \times 10^{-7}$ \\
1985A & 26.8  & 55   & 20  & $1.2 \times 10^{26}$ & $2.5 \times 10^{-7}$ \\
1985B & 28.0  & 69   & 20  & $3.1 \times 10^{26}$ & $6.1 \times 10^{-7}$ \\
1986A & 46.1  & 57   & 6   & $2.6 \times 10^{26}$ & $9.2 \times 10^{-7}$ \\
1986G & 5.5   & 28   & 6   & $5.0 \times 10^{25}$ & $1.7 \times 10^{-7}$ \\
1986O & 28    & 71   & 6   & $1.3 \times 10^{26}$ & $7.4 \times 10^{-7}$ \\
1987D & 30    & 83   & 6   & $1.3 \times 10^{26}$ & $8.4 \times 10^{-7}$ \\
1987N & 37.0  & 67   & 20  & $4.2 \times 10^{26}$ & $7.4 \times 10^{-7}$ \\
1989B & 11.1  & 15   & 3.6 & $8.1 \times 10^{24}$ & $3.3 \times 10^{-8}$ \\
1989M & 17.4  & 50   & 6   & $9.2 \times 10^{25}$ & $4.4 \times 10^{-7}$ \\
1990M & 39.4  & 32   & 3.6 & $1.5 \times 10^{26}$ & $5.4 \times 10^{-7}$ \\
1991T & 14.1  & 28   & 3.6 & $2.3 \times 10^{25}$ & $1.5 \times 10^{-7}$ \\
1991bg& 17.4  & 29   & 3.6 & $1.1 \times 10^{26}$ & $2.0 \times 10^{-7}$ \\
1992A & 24.0  & 29   &  6  & $4.1 \times 10^{25}$ & $1.6 \times 10^{-7}$ \\
1994D & 14    & 61   & 6   & $2.8 \times 10^{25}$ & $2.5 \times 10^{-7}$ \\
1995al& 30    & 17   & 20  & $1.7 \times 10^{26}$ & $1.2 \times 10^{-7}$ \\
1996X & 30   & 66   & 3.6 & $1.9 \times 10^{26}$ & $1.2 \times 10^{-6}$ \\
1998bu& 11.8  & 28   & 3.6 & $1.3 \times 10^{25}$ & $1.1 \times 10^{-7}$ \\
1999by& 11.3  & 15   & 3.6 & $2.1 \times 10^{25}$ & $8.0 \times 10^{-8}$ \\
2002bo& 22    & 95   & 20  & $6.8 \times 10^{25}$ & $3.0 \times 10^{-7}$ \\
2002cv & 22 & 41 & 20 &  $6.8 \times 10^{25}$ & $3.0 \times 10^{-7}$ \\
2003hv & 23   & 61   & 3.6 & $6.2 \times 10^{25}$ & $5.8 \times 10^{-7}$ \\
2003if & 26.4 & 68   & 3.6 & $8.1 \times 10^{25}$ & $7.6 \times 10^{-7}$ \\
\enddata
\tablenotetext{a}{The spectral luminosity upper limit ($2\sigma$), as
estimated at the wavelength given in column (4), which, when combined
with the age of the SN at the time of observation, yielded the lowest
mass-loss rate limit. }
\tablenotetext{b}{The upper limit ($2\sigma$) to the mass-loss rate,
$\dot M$, is calculated from the spectral luminosity lowest upper limit
given in column (5), as measured at the wavelength given in column (4)
at an epoch after explosion given in column (3).   The mass-loss limits
are calculated with the assumption that the SN Ia progenitor systems can
be modeled by the known properties of SNe Ib/c progenitor systems, and
that the pre-SN wind velocity establishing the CSM is $w_{\rm wind} =
10$ \kms.}
\label{tab3}
\end{deluxetable}


\end{document}